\documentclass[12pt]{article}
\usepackage{amsmath}
\usepackage{amssymb}
\usepackage{bbm}
\usepackage{epsfig}
\usepackage{yfonts}

\def\R{{\mathcal R}}
\def\Q{{\mathcal Q}}
\def\N{{\mathcal N}}
\def\be{\begin{equation}}
\def\ee{\end{equation}}
\def\lp{\ell_P}

\def\a{\alpha}
\def\b{\beta}
\def\g{\gamma}
\def\d{\delta}
\def\e{\epsilon}
\def\f{\phi}
\def\fin{f_\infty}
\def\r{\rho}
\def\bnabla{\overline\nabla}
\def\bbox{\overline{\Box}}

\begin{document}

\title{Counterterms, critical gravity and holography}

\author{{ Kallol Sen}$^{1}$,
{ Aninda Sinha}$^{1,2}$ and {Nemani V. Suryanarayana}$^{3}$\\
\it $^1$Centre for High Energy Physics, Indian Institute of Science,  Bangalore, India.\\
\it $^2$Perimeter Institute for Theoretical Physics, Waterloo, Canada.\\
\it $^3$Institute of Mathematical Sciences, Chennai, India.}
\maketitle

\begin{abstract}{We consider counterterms for odd dimensional holographic CFTs. These counterterms are derived by demanding cut-off independence of the CFT partition function on $S^d$ and $S^1 \times S^{d-1}$. The same choice of counterterms leads to a cut-off independent Schwarzschild black hole entropy. When treated as independent actions, these counterterm actions resemble critical theories of gravity, i.e., higher curvature gravity theories where the additional massive spin-2 modes become massless. Equivalently, in the context of AdS/CFT, these are theories where at least one of the central charges associated with the trace anomaly vanishes. Connections between these theories and logarithmic CFTs are discussed. For a specific choice of parameters, the theories arising from counterterms are non-dynamical and resemble a DBI generalization of gravity. For even dimensional CFTs, analogous counterterms cancel log-independent cut-off dependence.}
\end{abstract}

\tableofcontents
\section{Introduction}

This paper considers apparently disparate topics : AdS counterterms and critical gravity.
Firstly, it is well known that in order to get finite results from AdS/CFT one needs to add counterterms to the bulk gravity lagrangian \cite{bk, robcounter, krauslarsen}. These counterterms are made from curvature invariants formed from the boundary metric and resemble a higher derivative gravity action in one lower dimension with the coefficients of the higher curvature terms chosen so that power law divergences in the bulk are canceled for all possible boundary topologies permitted by the equations of motion. These counterterms are crucial to get the area law for black holes in the Euclidean action approach. They are also important in connecting the Randall-Sundrum scenario \cite{rs} with AdS/CFT \cite{gub}. For odd $d$-dimensional CFTs, they are also crucial to get a finite free energy when the theory is placed on an $S^d$. In recent times, there has been a flurry of activity related to the conjecture \cite{mscarb, mch, klebaF} that the free energy for CFTs on $S^d$ in odd dimensions could satisfy a monotonicity property analogous to the famous Zamolodchikov's c-theorem in 1+1 dimensions.


Secondly, in recent times, it has become a widely pursued activity to consider higher curvature gravity theories with a certain choice of fixed coefficients for the higher derivative terms. Partly, interest in these theories started with the work of \cite{bht}, who found that in 3-dimensions, by fine tuning the parameters appearing in a gravity lagrangian including upto four derivative curvature terms, one can have a unitary, renormalizable theory around flat space. They in turn were motivated by the work of \cite{strom} who attempted to construct a chiral theory of gravity in 3-dimensions involving the Chern-Simons term. The way the ``new massive gravity" (NMG) of \cite{bht} worked was through the realization that in 3-dimensions, the usual graviton has no degrees of freedom and so one can flip the sign of the kinetic term, thereby also flipping the sign of the kinetic terms for the massive modes. This trick would certainly not work in higher dimensions so one can ask what other features of such theories could be reproduced in higher dimensions.

Four derivative theories in any dimensions could be tuned in a way to reach a ``critical point" where the massive spin-2 modes become massless and where the propagating degrees of freedom are just spin-2 fields \cite{lupope, desertekin, ohta}. Constructing such theories in four dimensions was initially motivated by the hope that such theories could be unitary and renormalizable\footnote{Renormalizability of four derivative theories in 3+1 dimensions was shown in \cite{stelle}.}. It turns out that there are solutions which are logarithmic in the radial coordinate which would spoil unitarity \cite{mohsen, berglog, gru}. One could impose boundary conditions\cite{maldac} to truncate these modes but the resulting theory appears to be contentless\cite{lupopemalda} at the critical point. While it is an interesting exercise to move off criticality and impose suitable boundary conditions to get a unitary theory, we will be less restrictive and consider the possibility of having AdS duals to log-CFTs, which are inherently non-unitary. In addition to the criticality feature, it was pointed out \cite{sinha1,sinhacos} that in the context of (A)dS/CFT, NMG theories arose by demanding the existence of a simple c-theorem. It turns out that the relation between the quadratic interactions, $R_{ab}R^{ab}$ and $R^2$ to construct a critical gravity theory in arbitrary dimensions, coincides precisely with the c-theorem constraint \cite{mscarb,ms1}. Thus it may also be hoped that such attempts to construct interesting higher derivative theories would admit simple holographic c-theorems.

In \cite{jatkarsinha} it was realized that there is a similarity between the NMG action and the counterterm action in AdS$_4$. It was shown that the Dirac-Born-Infeld generalization of NMG \cite{tekinnmg} can arise as a suitable counterterm in AdS$_4$. This counterterm led to cut-off independent results for the Euclidean onshell action as well as the AdS$_4$-Schwarzschild black hole entropy.
The 3-dimensional DBI-gravity action considered in \cite{jatkarsinha} was such that the mass of the additional spin-2 mode in NMG was pushed to zero. This is reminiscent of what happens in critical gravity! Furthermore, the central charge of the putative dual 1+1 dimensional CFT was also shown to be zero.
As we will show in this paper, critical theories of gravity have at least one central charge that vanishes. Motivated by these observations, we study counterterm actions in AdS in higher dimensions to see if resulting counterterm actions (taken as independent actions) are critical.
Schematically, the total action $I$ is written as
\be
I_{tot}=I^{d+1}_{bulk,M}+I^{d}_{GH,\partial M}+I^{d}_{ct,\partial M}\,,
\ee
where $I^{d+1}$ is the $d+1$-dimensional Einstein-Hilbert action, $I^{d}_{GH}$ is the standard Gibbons-Hawking term while $I^{d}_{ct}$ are the AdS counterterms, written in terms of the $d$-dimensional boundary metric. As noted above, the counterterms are needed to make the on-shell results, which contain physical information like the free energy on a sphere, black hole entropy etc, finite. In addition to demanding that the divergences cancel, we will derive counterterms such that the total action, and hence the free energy on a sphere ($\partial M=S^{d}$), black hole entropy ($\partial M=S^1\times S^{d-1}$) are cut-off independent. At zero temperature, there is a Casimir energy associated with $\partial M=S^1\times S^{d-1}$ \cite{bk,robcounter} for $d=$even which in turn is related to the trace anomaly. Since the trace anomaly does not depend on the cut-off, it is reasonable to expect that the Casimir energy and hence the on-shell Euclidean action is also independent of the cut-off. At finite temperatures, we will show that the cut-off independent black hole entropy is perfectly consistent with the first law of thermodynamics. We will take $I=I^{d}_{ct}$ as a ``stand-alone" action in $d$-dimensions and study its properties.

Our findings can be briefly summarized as follows: We find that there exist counterterms that cancel the cutoff dependence for odd dimensional CFT partition functions on $S^d$ and $S^1 \times S^{d-1}$. We explicitly show these counterterms for $d=3,5,7$. The same counterterms can be shown to lead to a Schwarzschild black hole entropy that is exactly $S=A/4$ without any cutoff dependence, if one considers a finite cut-off. Consistency with the first law of thermodynamics is explicitly checked. When treated as independent actions, the equations of motion for linearized fluctuations, after a field redefinition, are identical to what appears in critical gravity. The similarities and differences between existing critical gravity models and what arises from our actions are pointed out. We also comment on the role played by analogous counterterms in even dimensions in the discussion.


This paper is organized as follows: In section 2, we review the construction of quadratic critical gravity theories. We compute the CFT trace anomalies in 1+1 and 3+1 dimensions. In 1+1 dimensions, the Euler anomaly vanishes. We show that in the presence of the Gauss-Bonnet term the Euler anomaly does not vanish in 3+1 dimensions but the central charge $c$ vanishes. In section 3, we turn to the calculation of counterterms which lead to cutoff independent results for boundaries with vanishing Weyl tensor. In section 4, we calculate the entropy of Schwarzschild black hole using Euclidean methods and demanding a finite cutoff. We demonstrate that the first law of black hole thermodynamics is respected. In section 5, we consider these theories as independent actions and work out their properties. We demonstrate that the linearized equations of motion after a field redefinition, coincide precisely with those of critical gravity. In section 6, we argue that there is a logarithmic partner to the vacuum as dictated by the putative log-CFT dual. We conclude in section 7. There are four appendices containing useful details of the calculations that have gone into the main paper. In the fifth appendix, we address c-theorem constraints on the lagrangians.

\noindent{\bf Conventions}: We will write the Einstein Hilbert action as
\be
\frac{1}{2\lp^{n-2}}\int d^n x \sqrt{-g}(R-2\Lambda_0)\,,
\ee
and use $2\Lambda_0=-(n-1)(n-2)/L^2$ for AdS spaces. When we have a higher derivative theory of gravity, the AdS radius is typically dependent on the higher derivative terms. We will write the Ricci tensor as $R_{ab}=-(n-1)\fin/L^2 g_{ab}$. We will denote a perturbation by $h_{ab}$. The background values for various quantities will be indicated by a bar. For any AdS/CFT analysis involving bulk and counterterms, we will reserve $D$ for the bulk dimensions and $d=D-1$ for the boundary dimensions. For most of this paper $D$ will be even while $d$ will be odd.  AdS
metrics solving Einstein gravity can be written as

\begin{equation}\label{adsm}
ds^2=\frac{dr^2}{k+\frac{r^2}{L^2}}+(k+\frac{r^2}{L^2})
d\hat\Sigma^2_{-k,\hat m}+\frac{r^2}{L^2}d\tilde \Sigma^2_{k,\tilde
  m}\,,
\end{equation}
where $k=0,\pm 1$. The metrics $d\hat\Sigma^2_{-k,\hat m}$ and
$d\tilde \Sigma^2_{k,\tilde m}$ are defined through
\begin{equation}
  \label{eq:met}
d\Sigma^2_{k,m}=\begin{cases} L^2 d\Omega_m^2  & \mbox{for~} k=+1\\
                              \sum_{i=1}^m dx_i^2 & \mbox{for~} k=0 \\
                              L^2 d\Psi_m^2  & \mbox{for~} k=-1
                \end{cases}
\end{equation}
where $d\Omega_m^2$ is the metric on a unit $m$-sphere and $d\Psi_m^2$
is the metric on a $m$-hyperboloid whose metric is obtained by the
analytic continuation of the metric on the unit $m$-sphere. In higher derivative theories, $L^2$ is replaced by $L^2/\fin$. The cutoff is $r=\Lambda$.

\section{Review of quadratic critical gravity}
In this section we will review some salient features in the construction of critical gravity theories \cite{lupope,desertekin}.
\subsection{Construction of critical gravity}
Quadratic critical gravity is a special theory of gravity including up to four derivative curvature terms in addition to the Einstein-Hilbert term and a cosmological constant. We start with the action
\be
\frac{1}{2\lp^{n-2}}\int d^n x \sqrt{-g}\left[R-2\Lambda_0+\alpha L^2 R^2+\beta L^2 R_{ab}R^{ab}+\gamma L^2 GB \right]\,,
\ee
where $GB=(R_{abcd}R^{abcd}-4 R_{ab}R^{ab}+R^2)$ is the Gauss-Bonnet term.
Writing the Ricci tensor as $R_{ab}=2\fin \Lambda_0/(n-2)$ leads to the equations of motion giving
\be\label{eqfin}
1-\fin+[\frac{(n-4)(n-1)(n\alpha+\beta)}{n-2}+(n-3)(n-4)\gamma]\fin^2=0\,.
\ee
Typically there are 2 vacua and one needs to either arbitrarily consider one of them or specify some further conditions which leads to a unique vacuum. The addition of higher curvature terms generically leads to additional massive spin 2 and spin 0 modes. This is seen by considering fluctuations around an AdS vacuum, $g_{ab}=\bar g_{ab}+h_{ab}$. Diffeomorphisms allow us to choose the gauge condition $\bnabla_a h^{ab}=\bnabla^b h$ where $h=h^a_a$, the trace. The equations of motion lead to
\begin{eqnarray}
\left[H_1(\alpha,\beta)\bbox+\frac{H_2(\alpha,\beta,\g)}{L^2}\right]h&=& 0\,,\\
\beta\left[\bbox+\frac{2\fin}{L^2}-M(\alpha,\beta,\g)^2\right]\left[\bbox+\frac{2\fin}{L^2}\right]h_{ab}&=& 0\,.
\end{eqnarray}
At this stage, there appear to be propagating scalar and spin-2 ghosts. For our purpose it is sufficient to note that the conditions for $H_1,H_2, M$ to be zero are given by:
\begin{eqnarray}
H_1=0 &\implies& \alpha= -\frac{n\beta}{4(n-1)}\,,\label{ctobs}\\
H_2=0 &\implies& 1-2\fin[\frac{(n-1)(n-4)}{n-2}(n\alpha+\beta)+(n-3)(n-4)\gamma]= 0\,,\label{h2zero} \\
M=0 &\implies& 1-2\fin[(n-1)(n\alpha+\beta)+(n-3)(n-4)\gamma]= 0\,.
\end{eqnarray}
We are interested in $M=0$ \cite{lupope}. Consider first $n=3$. It is easily seen that $H_2=0$ is not compatible with $M=0$ unless $\fin=0$. Thus $H_1=0$ is imposed by tuning $\alpha,\beta$. For $n=4$ again, we see that we cannot set $H_2=0$. In higher dimensions $n>4$, there appears to be a choice: Either we set $H_1=0$ with $H_2\neq 0$ so that there is no propagating scalar $h$; or we set $H_2=0$ with $H_1\neq 0$ so that there is a propagating massless scalar. Of course, one has to check if the resulting choice leads to sensible physical parameters or not. What is usually done is to set $H_1=0$ with $H_2\neq 0$ following what happens in $n=3,4$. For $n=5$ if we set $H_2=0, H_1\neq 0$, we would get $\alpha=-\beta/5, \gamma=1/8, \fin=2$ with $H_1=9\beta/5$. As we will show in the next section, the Euler anomaly of the dual CFT becomes negative in this case indicating an inherent sickness of the underlying theory. One can point out that the positivity of the Euler anomaly is a condition arising from unitarity and since the underlying theory is related to log-CFTs which are not unitary, why should we care? However, as we alluded to above, in all theories considered so far, there is a possibility to consider specific boundary conditions which truncate the non-unitary modes. With a negative Euler anomaly, this would be impossible. Hence for the quadratic theories, we will consider  $H_1=0$ with $H_2\neq 0$.
This is what is called a ``critical-point" in the literature and the resulting theory is called critical gravity\footnote{Since the propagator has changed from $1/p^2(p^2+m^2)\sim 1/m^2(1/p^2-1/(p^2+m^2))$ to $1/p^4$, it may have been hoped that the ghost problem of the untuned theory would be ameliorated. This expectation is supported by the analysis\cite{bender} of the Pais-Uhlenbeck oscillator at the critical point using methods of PT quantum mechanics which claims to have gotten rid of the unwanted ghosts. Unfortunately, the link with logarithmic CFTs seems not to support this point of view. It may be fruitful to repeat the path integral analysis of Hawking and Hertog\cite{hh} at the critical point to see how severe the problem with ghosts is in a cosmological scenario.}. The spin-2 fields satisfy
\be
\left[\bbox+\frac{2\fin}{L^2}\right]^2 h_{ab}=0\,.
\ee
This admits solutions of the form
\be
\left[\bbox+\frac{2\fin}{L^2}\right]h^E_{ab}=0\,,
\ee
which are the usual Einstein modes as well as
\be
\left[\bbox+\frac{2\fin}{L^2}\right]h^L_{ab}\propto h^E_{ab}\implies \left[\bbox+\frac{2\fin}{L^2}\right]^2 h^L_{ab}=0\,.
\ee
$h^L_{ab}$ are the so-called logarithmic modes. Evidently, the operator $\bbox$ has a matrix representation
\be
\bbox\rightarrow
\begin{pmatrix}
-\frac{2\fin}{L^2} & \# \\
0 & -\frac{2\fin}{L^2}
\end{pmatrix}
\ee
on the space $(h^L, h^E)$ and is not Hermitian. $(h^L, h^E)$ form a rank-2 Jordan cell. Rank-$n$ Jordan cells are the starting point for defining logarithmic conformal field theories. The states $h^E$ turn out to be null states and in order to construct a unitary Hilbert space\footnote{Of course, this uses a particular definition of the norm of states.}, they should be modded out \cite{porrati}. Since in $n>3$, $h^E$ and $h^L$ mix, modding out $h^E$ would in turn mod out the $h^L$ states as well. Thus we would be left with a trivial vacuum. Thus these theories seem to be necessarily non-unitary. Although the original motivation for studying these critical theories seem to have dampened, their connection with logarithmic CFTs make them appealing nonetheless to probe further.

Another viewpoint that has been advocated by Maldacena \cite{maldac} is that the log modes can be got rid of by imposing suitable boundary conditions. The truncated theory in this case would  simply coincide with Einstein gravity. This point of view has been studied for example in \cite{lupopemalda} in the context of critical gravity. Throwing away the log modes leads to a contentless theory as the massless gravitons have zero energy. This has led to consideration of theories away from criticality but where non-unitary modes can be truncated using suitable boundary conditions. At this stage, it is not entirely clear if such a procedure makes sense at a quantum level where loops are involved. 
For the purpose of the present work, it is useful to keep in mind both points of view.

\subsection{Anomalies in dual CFTs}

In 1+1 dimensions, unitary conformal field theories are  specified by the central charge $c$. When $c=0$, one has new logarithmic operators needed to have a well defined operator product expansions \cite{cardylog}. In 3+1 dimensions CFTs are specified by two central charges $c$ and $a$. So one question naturally arises as to which central charge should be set to zero in order to have a log-CFT. Since
$$\langle T_{ab} T_{cd} \rangle\propto c \langle 0|0\rangle =0\,,$$
there are three possibilities \cite{koganclass}:
\begin{enumerate}
\item $c=0 , \langle 0|0\rangle \neq 0$. This corresponds to having a unique vacuum.
\item $c=0 , \langle 0|0\rangle=0$. This corresponds to have a log degenerate vacuum as well as a vanishing central charge.
\item $c\neq 0, \langle 0|0\rangle = 0$. This corresponds to the case where the vacuum has a logarithmic partner.
\end{enumerate}
To extract the Euler anomaly ($c$ in 1+1 dimensions, $a$ in 3+1 dimensions), we put the CFT on a $S^{n-1}$ and extract the logarithmically divergent term. After some algebra, the onshell action at criticality (without imposing eq.(\ref{ctobs}) or eq.(\ref{h2zero})) works out to be
\be
\frac{4}{\lp^{n-2}}V_{S^{n-1}}\frac{\fin^2}{L^2}\gamma (n-1)(n-3)^2  \int dr\, r^{n-1}\sqrt{g_{rr}} \,.
\ee
For $n=3$ or when the putative dual CFT is in 1+1 dimensions, we thus find that $c=0$ at the critical point.
For $n=5$, using the method described in appendix (\ref{simpan}), we compute the Euler anomaly $a$ and the Weyl anomaly $c$ to find out which of the above possibilities is realized. We find that at the critical point
\be
c=0\,, \quad a=-\frac{8\pi^2 L^3 }{\fin^{1/2}\lp^3 }\gamma\,.
\ee
Thus in the absence of the Gauss-Bonnet term both anomalies would vanish. It is interesting to note that in order to have $a>0$ we would need $\gamma<0$. The fact that black hole entropy in the critical theories vanish although $a\neq 0$ is confusing at first in light of  the $a$-theorem which proposes that $a$ counts the number of degrees of freedom in the CFT \cite{mscarb}. However, considering the fact that the critical theories are not unitary, this is probably not surprising and is not a contradiction with \cite{mscarb}. Thus these theories realize either (1) or (2) of the possibilities listed above. We will find out later that there exist a whole class of theories where (3) is realized.

\subsection{Similarities with counterterms}
We now point out a similarity of the critical gravity action with AdS counterterms. The fact that new massive gravity or its Dirac-Born-Infeld extension has something to do with AdS counterterms was already pointed out in \cite{jatkarsinha}.
It is well known that in order to make sense of the bulk action in the
AdS/CFT correspondence, one needs to add counterterms since the
gravity action typically diverges
\cite{bk,robcounter,krauslarsen,sken}.  In \cite{robcounter} for
example, it was shown that the full (Euclidean) gravitational action
in $D=d+1$ spacetime dimensions has three contributions
\begin{equation}
  \label{eq:actads}
I_{AdS}=I_{bulk}(g_{\alpha\beta})+I_{surf}(g_{\alpha\beta})+I_{ct}(\gamma_{mn})\,,
\end{equation}
where $I_{bulk}$ is the familiar classical action given by
\begin{equation}
  \label{eq:actblk}
I_{bulk}=-\frac{1}{2\ell_P^{d-1}} \int_{{\mathcal M}} d^{d+1} x
\sqrt{g}\left(R+\frac{d(d-1)}{L^2}\right)\,,
\end{equation}
$I_{surf}$ is the Gibbons-Hawking term given by
\begin{equation}
  \label{eq:surf}
I_{surf}=-\frac{1}{\ell_P^{d-1}}\int_{\partial {\mathcal M}} d^d x \sqrt{\gamma} K\,,
\end{equation}
with $K=\gamma^{mn}\nabla_m \hat n_n$ being the trace of the extrinsic
curvature of the boundary.  Here $\g_{mn}$ is the induced metric on
the boundary defined through $\g_{mn}=g_{mn}-\hat n_m \hat n_n$ with
$\hat n$ being an outward pointing unit normal vector to the boundary. The
counterterm action $I_{ct}$ can be arranged as an expansion in powers
of the boundary curvature \cite{robcounter}:
\begin{eqnarray}\label{robeq}
I_{ct}&=&\frac{1}{\ell_P^{d-1}}\int_{\partial {\mathcal M}} d^d
x\sqrt{\g}\left[\frac{d-1}{L}+\frac{L}{2(d-2)} \R \right. \nonumber\\
&& ~~~~~~~~~\left.+\frac{L^3}{2(d-4)(d-2)^2}\left(\R_{ab}\R^{ab}-
    \frac{d}{4(d-1)}\R^2\right)+\cdots\right]\,,
\end{eqnarray}
where $\R$ and $\R_{ab}$ are the Ricci scalar and Ricci tensor made
out of $\g_{ab}$. The relative coefficients between the quadratic terms is precisely what is needed to eliminate the scalar mode as in eq.(\ref{ctobs}) in any dimensions. This gives us motivation to look for similarities between the counterterm actions and critical gravity. Firstly we note that in order to get $c=0$ in the putative 1+1 CFT duals, it was necessary to consider the DBI completion of the counterterm in \cite{jatkarsinha}. The truncated stand-alone quadratic theory in eq.(\ref{robeq}) is not a critical theory. The DBI completion arose by demanding that the Euclidean onshell action was cutoff independent to {\it all orders}. The DBI form of the counterterm was guessed in \cite{jatkarsinha}. Below we will show that there is a one parameter extension of this action which has $c=0$ and whose linearized equations of motion are what arise in critical gravity. Then we will go onto generalize this construction in higher dimensions.

\section{Counterterm actions}
In this section we will present counterterm actions that make the Euclidean on shell action cutoff independent to all orders for boundaries having topology $R \times S^{d-1}$ and $S^d$. Boundary topologies of the form $R\times H_{d-1}$, $H_d$ follow trivially. The Weyl tensor for these boundaries vanishes. We will comment on boundaries of the form $H_p\times S^q$ separately since the Weyl tensor is generally nonvanishing in these cases\footnote{While it is true that when the cut-off is taken to infinity, the boundaries with topology $H_p\times S^q$ are conformally flat, this is no longer true at a finite cut-off.}. We will deal with conformally flat boundaries. Our counterterm action will enable the calculation of correlation functions of the stress tensor in the context of holographic renormalization group flows. Furthermore, our counterterm actions lead to a cutoff independent black hole entropy when the black hole entropy is calculated using Euclidean methods. We begin with the results:\footnote{This result is valid for odd $p+q$.}
\begin{multline} \label{ctr}
\frac{(I_{bulk} + I_{surf})|_{\partial M = H_p \times S^q}}{(\frac{L}{l_p})^{p+q-1} V_{H_p} V_{S^q}} = \frac{p+q}{1+q} \Big(\frac{\Lambda}{L}\Big)^{q+1} {}_2F_1 \left[\frac{1-p}{2}, \frac{1+q}{2}, \frac{3+q}{2}, - \frac{\Lambda^2}{L^2} \right] \\
 - \Big(\frac{\Lambda}{L}\Big)^{q-1} \Big( 1+ \frac{\Lambda^2}{L^2} \Big)^{\frac{p-1}{2}}\Big[p \, \frac{\Lambda^2}{L^2} + q \,\Big(1+ \frac{\Lambda^2}{L^2} \Big) \Big]
 + \frac{\Gamma[1- \frac{p+q}{2}] \Gamma[\frac{1+q}{2}]}{\Gamma[\frac{1-p}{2}]}
\end{multline}
Here for $p+q$ odd, the first term has the form $\displaystyle\sum_{n=0}^{1}\sum_{m}c_{n,m}(1+\Lambda^2/L^2)^{n/2}(\Lambda^2/L^2)^m$.
%
Let us define the following
\be
\R_n={\rm tr}\,(\R^n)\,,
\ee
or explicitly
\be
\R_2= \R_a^b \R_b^a\,,\quad
\R_3= \R_a^b \R_b^c \R_c^a\,,\quad {\rm etc}
\ee
At $O(\R^n)$ there are $p(n)$ number of independent terms where $p(n)$ is the number of ways to partition $n$. The onshell results above can be written as $\sqrt{f(\Lambda)}$ and since the minimum power of the cutoff in $f(\Lambda)$ is $\Lambda^0$, while $\R\sim 1/\Lambda^2$ with $\sqrt{h}\sim \Lambda^d$, it is easy to see that if the counterterm was written as $\sqrt{h}\sqrt{1+c_i O(\R^i)}$ then the maximum power of $\R$ involved would be at most $d$. Thus we start with $c_d=1+\sum_{n=1}^{n=d} p(n)$ parameters in the ansatz for the counterterms. Furthermore, we see by expanding eq.(\ref{ctr}), that at each order in $\Lambda$, there will be two relations. So we will be left with a $c_d-2 d$ parameter counterterm. For example, when $d=3$ we should get a 1 parameter solution, for $d=5$ we should get a 9 parameter solution and for $d=7$ a 31 parameter solution. The ambiguity arises since there are precisely $c_d-2$ independent invariants made of $\R_a^b$ which vanish for both $S^d$ and $R\times S^{d-1}$. We have checked this for various cases but we do not know of a general proof of this statement.
\subsection{$D=4/d=3$}
\begin{eqnarray}
I^{(3)}_{ct}&=&\frac{2}{L \lp^2}\int d^3 x \sqrt{\g}\bigg{(}1+\frac{1}{2} L^2 \R-\frac{1}{2}L^4(\R_2-\frac{1}{2}\R^2)\nonumber \\&&~~~~~~~~~~~~~~~~+\lambda L^6 \R^3+(\frac{1}{24}-5\lambda)L^6 \R\R_2+(6\lambda-\frac{1}{12})L^6 \R_3\bigg{)}^{1/2}\,.
\end{eqnarray}
Thus we have a one parameter solution. For the choice $\lambda=-1/24$ the action coincides with the DBI action in \cite{jatkarsinha}.
\subsection{$D=6/d=5$}
\begin{eqnarray}
I^{(5)}_{ct}=\frac{4 }{L\lp^4}\int d^5 x \sqrt{\g}\bigg{(} &1&+\frac{1}{12} L^2 \R+\frac{1}{36}L^4 (\R_2-\frac{1}{4} \R^2)\nonumber \\
&+& L^6(\b_5 \R_3+\b_6 \R_2 \R+\b_7 \R^3)\nonumber \\
&+& L^8(\g_8 \R_4+\g_9 \R_3 \R+\g_{10} \R_2^2+\g_{11} \R_2 \R^2+\g_{12}\R^4)\nonumber \\
&+& L^{10}(\d_{13} \R_5+\d_{14} \R_4 \R+\d_{15}\R_3 \R_2+\d_{16}\R_3 \R^2 \nonumber \\&~&~~~~~~~~~~~~~~+\d_{17}\R_2 \R^3+\d_{18} \R_2^2 \R+\d_{19} \R^5)\bigg{)}^{1/2}\,,
\end{eqnarray}
with
\begin{eqnarray}\label{beta5}
\b_5 &=& \frac{1}{48}+20\b_7\,,\quad \b_6=-\frac{1}{192}-9\b_7\,,\\
\g_8 &=& 20(\g_{11}+9\g_{12}) \,,\quad \g_9=-\g_{10}-9\g_{11}-61 \g_{12}\,,\\
\d_{13}&=& -\frac{1}{2880}+20(\d_{16}+9\d_{17}+\d_{18}+61 \d_{19})\,,\\
\d_{14} &=& \frac{1}{11520}-(\d_{15}+9\d_{16}+61 \d_{17}+9\d_{18}+369 \d_{19})\,.
\end{eqnarray}
Thus we are left with a 9 parameter solution.

\subsection{$D=8/d=7$}

\begin{eqnarray}
I_{ct}^{(7)} = \frac{6}{L \, l_p^6} \int d^7x \, \sqrt{\gamma} \, \bigg{(}1 &+& \frac{1}{30} L^2 \, {\cal R} + \frac{1}{450} L^4 \, ({\cal R}_2 - \frac{1}{6} {\cal R}^2) \cr
&+& L^6 (\b_5 \R_3 + \b_6 \R_2 \R + \b_7 \R^3)\nonumber \\
&+& L^8 (\g_8 \R_4+\g_9 \R_3 \R+\g_{10} \R_2^2+\g_{11} \R_2 \R^2+\g_{12}\R^4)\nonumber \\
&+& L^{10} (\d_{13} \R_5+\d_{14} \R_4 \R + \d_{15} \R_3 \R_2+\d_{16}\R_3 \R^2 +\d_{17} \R_2 \R^3 \nonumber\\~~~~~~~~~~~~~~~~~~~&+&\d_{18}\R_2^2 \R+\d_{19} \R^5)\nonumber \\
&+&L^{12}(\e_{20}\R_{6}+\e_{21}\R_{5}\R+\e_{22}\R_{4}\R_{2}+\e_{23}\R_{4}\R^{2}+\e_{24}\R_{3}^{2}\nonumber\\
&+&\e_{25}\R\R_{2}\R_{3}+\e_{26}\R^{3}\R_{3}+\e_{27}\R_{2}^{3}+\e_{28}\R_{2}^{2} \R^{2} +\e_{29}\R_{2}\R^{4}+\e_{30}\R^{6})\nonumber\\
&+&L^{14}(\f_{31}\R_{7}+\f_{32}\R_{6}\R+\f_{33}\R_{5} \R_{2}+\f_{34}\R_{5}\R^{2}
+\f_{35}\R_{3}\R_{4}\nonumber \\&+& \f_{36} \R \R_{2} \R_{4}+\f_{37} \R^{3}\R_{4}
+\f_{38}\R\R_{3}^{2} + \f_{39} \R_{2}^{2} \R_{3} \nonumber \\&+&\f_{40}\R^{2}\R_{2} \R_{3}
+\f_{41}\R^{4}\R_{3}+\f_{42}\R \R_{2}^{3}+\f_{43}\R^{3} \R_{2}^{2} \nonumber \\ &&~~~~~~~~~~~~~~~~~~+\f_{44}\R^{5}\R_{2}+\f_{45}\R^{7})\bigg{)}^{1/2}\nonumber \\
\end{eqnarray}
\begin{eqnarray}
\beta_5 &=& - \frac{1}{540} + 42 \, \beta_7, ~~ \beta_6 = \frac{1}{3240} - 13 \, \beta_7 \cr
\gamma_8 &=& - \frac{1}{1215} + 42 \, \gamma_{11} + 546 \, \gamma_{12}, ~~
\gamma_9 = \frac{1}{7290} - \gamma_{10} - 13 \, \gamma_{11} - 127 \, \gamma_{12}, \cr
\d_{13}&=&42(\d_{16}+13 \d_{17}+\d_{18}+127 \d_{19})\,,\\
\d_{14}&=&-\d_{15}-13\d_{16}-127 \d_{17}-13\d_{18}-1105\d_{19}\,,\\
\epsilon_{20} &=& 42 \, (\e_{23} + \e_{25} + 13 \, \e_{26} + \e_{27} + 13 \, \e_{28} + 127 \, \e_{29} + 1105 \, \e_{30}), \cr
\e_{21} &=& - \e_{22} - 13 \, \e_{23} - \e_{24} - 13 \, \e_{25} -127 \, \e_{26} -13 \, \e_{27} - 127 \, \e_{28} - 1105 \, \e_{29} - 9031 \, \e_{30} , \cr
\f_{31} &=& - \frac{1}{1148175} + 42 \, (\f_{34} + \f_{36} + \f_{38} + \f_{39}) + 546 \, (\f_{37} + \f_{40} + \f_{42}) \cr
&& ~~~~~~~~~~~~~~~~ + 5334 \, (\f_{41} + \f_{43}) + 46410 \, \f_{44} + 379302 \, \f_{45} , \cr
\f_{32} &=& \frac{1}{6889050} - (\f_{33} + \f_{35}) - 13 \, (\f_{34} + \f_{36} + \f_{38} + \f_{39}) - 127 \, (\f_{37} + \f_{40} + \f_{42}) \cr
&& ~~~~~~~~~~~~~~~~~~~~~~~ - 1105 \, (\f_{41} + \f_{43}) - 9031 \, \f_{44} - 70993 \, \f_{45} \, .
\end{eqnarray}
Thus we have a 31 parameter solution.

\subsection{Weyl corrections}
In the previous subsections we presented the counterterm actions required to make the Euclidean actions of AdS gravity theories cutoff independent in locally AdS backgrounds with conformally flat boundary metrics. However, the induced metric on the boundary of locally AdS spaces are not necessarily conformally flat. For such geometries the counterterm actions have to depend on the other components of the Riemann tensor of the boundary metric apart from the Ricci tensor. This dependence can be incorporated into the action through dependence on the Weyl tensor of the boundary metric.

Here we will demonstrate that by incorporating certain invariants involving the Weyl tensor we can indeed obtain the counterterm action for the $D=6$ ($5$-dimensional boundary). It can be verified that the following counterterm action is a possible answer for this case which makes the onshell action cutoff independent for all boundary topologies.
\begin{eqnarray}
I^{(5)}_{ct}=\frac{4 }{L\lp^4}\int d^5 x \sqrt{\g}\bigg{(} &1&+\frac{1}{12} L^2 \R+\frac{1}{36}L^4 (\R_2-\frac{1}{4} \R^2)\nonumber \\
&+& L^6(\b_5 \R_3+\b_6 \R_2 \R+\b_7 \R^3)\nonumber \\
&+& L^8(\g_8 \R_4+\g_9 \R_3 \R+\g_{10} \R_2^2+\g_{11} \R_2 \R^2+\g_{12}\R^4)\cr
&+& L^{10}(\d_{13} \R_5+\d_{14} \R_4 \R+\d_{15}\R_3 \R_2+\d_{16}\R_3 \R^2 \nonumber \\&~&~~~~~~~~~~~~~~+\d_{17}\R_2^2 \R +\d_{18}\R_2 \R^3 +\d_{19} \R^5)
\cr
&+& {\cal W}^2 \Big[ \alpha'_1 + \alpha'_2 L^2 \R +  L^4 (\alpha'_3 \R_2 + \alpha'_4 \R^{2}) \cr
&& ~~~~~~~~~ + L^6(\beta'_5 \R_3+\beta'_6 \R_2 \R+\beta'_7 \R^3) \Big]  \bigg{)}^{1/2}\,,
\end{eqnarray}
where
\begin{eqnarray}
\alpha'_1 &=& - \frac{11}{96} - 54 (\beta_7 +\gamma_{10} + 4 \gamma_{11} + 36 \gamma_{12}), \cr
\alpha'_2 &=& \frac{11}{1440} - \frac{9}{2} (\beta_7 +\gamma_{10} + 7 \gamma_{11} + 75 \gamma_{12} + \delta_{15})
-54 (\delta_{16} + 2 \delta_{17} + 13 \delta_{18} + 97 \delta_{19}), \cr
\alpha'_3 &=& \frac{5}{6912} + \frac{1}{10} (\beta_7 + \gamma_{10} - 11 \gamma_{11} - 59 \gamma_{12} - 15 \delta_{15}) - (21 \delta_{16} + 17 \delta_{17} + 193 \delta_{18} + 1397 \delta_{19}) \cr
\alpha'_4 &=& \frac{19}{34560} + \frac{1}{10} (\beta_7 + \gamma_{10} - 11 \gamma_{11} - 179 \gamma_{12} - 15 \delta_{15}) - (3 \delta_{16} + 11 \delta_{17} + 55 \delta_{18} + 419 \delta_{19}), \cr
\beta'_5 &=& \frac{1}{1280} + \frac{1}{10} (3 \beta_7 + 3 \gamma_{10} - 3 \gamma_{11} - 87 \gamma_{12} - 15 \delta_{15}) - (\delta_{16} + 7 \delta_{17} + 3 \delta_{18} - 113 \delta_{19}), \cr
\beta'_6 &=& - \frac{11}{17280} - \frac{1}{4} (\beta_7 + \gamma_{10} - \gamma_{11} - 29 \gamma_{12} - 5 \delta_{15}) - (\delta_{16} - 4 \delta_{17} + 15 \delta_{18} + 209 \delta_{19}), \cr
\beta'_7 &=& \frac{1}{6912} + \frac{1}{20} ( \beta_7 + \gamma_{10} - \gamma_{11} - 29 \gamma_{12} - 5 \delta_{15}) - \delta_{17} + 14 \delta_{19} \,
\end{eqnarray}
and $\beta_5$, $\beta_6$, $\gamma_8$, $\gamma_9$, $\delta_{13}$, $\delta_{14}$ are given by eq.(\ref{beta5}).

Note that this action is not the most general counterterm action we could have written down with the Weyl tensor included as there are many more invariants that can be constructed out of the Weyl tensor other than ${\cal W}^2$. This action, however, does not introduce any more free parameters than the conformally flat case presented earlier. In the rest of the paper we will not consider Weyl modifications.

\section{Black hole entropy}
The above choice of counterterms leads to a self-consistent way of producing the area law for Schwarzschild black holes{\footnote{We expect the Kerr black hole to also work but we have not been able to see how to define the cutoff surface at finite $\Lambda$. For charged black holes, we will need to add further terms involving $F_{ab}$.}}. The Euclidean method of computing black hole entropy requires knowledge of the counterterms. If we wanted to work with a counterterm action that just got rid of the divergences then we would get
$$
S_{BH}=\frac{A}{4}+O(1/\Lambda)\,,
$$
where $r=\Lambda$ is the cutoff and is taken to infinity. However if we did not take $\Lambda$ to infinity there is a possibility of getting $O(1/\Lambda)$ corrections. At first sight this seems at odds with the Wald entropy formalism which instructs us to evaluate the functional derivative of the lagrangian with respect to certain components of the Riemann tensor {\it at the horizon} and hence would not be dependent on the cutoff. Hence it is reasonable to assume that the correct choice of counterterms would lead to $S_{BH}=A/4$ with no cutoff dependence. Our choice of counterterms does precisely this \cite{jatkarsinha}. Let us demonstrate that this is also consistent with the first law of thermodynamics.
The boundary stress tensor \cite{bk} is defined as
\be
T_{\mu\nu}=\frac{1}{\lp^{D-2}}\left[K_{\mu\nu}-K\gamma_{\mu\nu}+\frac{2}{\sqrt{-\gamma}}\frac{\delta I_{ct}}{\delta \gamma^{\mu\nu}}\right]\,,
\ee
where the calculation of $\frac{\delta I_{ct}}{\delta \gamma^{\mu\nu}}$ has been done in the appendix. We write the boundary metric, making manifest a spacelike surface $\Sigma$, in an ADM manner as
\be
\gamma_{\mu\nu}dx^\mu dx^\nu=-N_\Sigma^2 dt^2+\sigma_{ab}(dx^a+N_\Sigma^a dt)(dx^b+N_\Sigma^b dt)\,,
\ee
and define the energy density as $\epsilon=u^\mu u^\nu T_{\mu\nu}$ where $u^\mu$ is a timelike unit normal to $\Sigma$.
From the stress tensor we define the mass as
\be
M=\int_\Sigma d^{D-2} x \sqrt{\sigma}N_\Sigma \epsilon\,.
\ee
Writing the Schwarzschild metric as
\be
ds^2=-[1+f(r)]N dt^2+\frac{dr^2}{1+f(r)}+g(r)d\Omega_{D-2}^2\,,
\ee
where $N=g(\Lambda)/(1+f(\Lambda))$ is chosen to ensure that the speed of light in the dual CFT is unity. This choice of $N$ turns out to be consistent with the first law of thermodynamics as well. We obtain
\be
M=\int d\Omega_{D-2} T_{tt} r^{(D-3)/2}\,.
\ee
For this metric the temperature works out to be
\be
T=\frac{\sqrt{N}f'(r_0)}{4\pi}\,,
\ee
where $r_0$ is the location of the horizon.
Using $S=A/4G=2\pi A/\lp^{D-2}$, we can check that
\be
\frac{\partial S}{\partial M}=\frac{1}{T}\,,
\ee
which is consistent with the first law of thermodynamics. Note here that the mass is now a function of the cutoff. In units of temperature, this is a monotonically increasing function of the cutoff. Using the knowledge that the entropy density can be written as $s=c_S T^{D-2}$ which is independent of the choice of the cutoff, we easily find $c_S(\Lambda)=\frac{c_S(\infty)}{N^{(D-2)/2}}$. $N$ exhibits a turning point close to the horizon after which it is a monotonically increasing function. So $c_S(UV)>c_S(IR)$ does not hold in general \cite{mscarb}. It will be interesting to use our counterterms to calculate higher point correlation functions of the stress tensor. We leave this for future work.

\section{Properties and criticality of counterterm actions}
In what follows we will consider the counterterm actions as stand-alone actions. Namely we will start with $I=I_{ct}$ and work out the equations of motion for the metric arising from this action. There are several key features of these actions:
\begin{enumerate}
\item There is a unique AdS vacuum. Unlike truncated theories, for example the quadratic critical theories, which have multiple (A)dS vacua, the square-root theories have a unique AdS vacuum.
\item The AdS vacuum has the same radius of curvature as what would have followed from the Einstein Hilbert action, namely
\be\label{ein}
I_{EH}=\frac{1}{2\lp^{n-2}}\int d^{n} x \sqrt{|g|}(R+\frac{(n-1)(n-2)}{L^2})\,.
\ee
\item The onshell action for the AdS vacuum is zero.
\item As shown in appendix E, for a specific choice of parameters, there exists a c-function \cite{sinha1,mscarb}.
\item The Euler anomaly is zero. To see this we consider the boundary $S^{n-1}$ and find that the on-shell action is identically zero.
\item The Weyl anomaly, $c$ for $n=5$ is not necessarily zero unless we impose a further relation on the parameters.
\item Solutions of eq.(\ref{ein}) are solutions of the square-root actions.
\item The entropy of the Kerr black hole is zero.
\item As we will show, the linearized spectrum around AdS is the same as that of a critical theory.
\item As we will later argue, there are AdS waves with logarithmic fall off with vanishing action that solve the equations of motion as well.
\end{enumerate}
Several of the above properties mimic quadratic critical gravity. In $n=5$, the key difference is that unlike quadratic critical gravity where $c$ is always vanishing, the square-root theories have vanishing Euler anomaly $a$ while $c$ is not necessarily zero. In order to have $c=0$ we need to have
either
\begin{eqnarray}
\b_7\equiv\b_c&=& -\frac{1}{960}-\frac{1}{3}(4\g_{10}-8\g_{11}-132\g_{12}-32\d_{15}+16\d_{16}+304\d_{17}\nonumber \\ &&~~~~~~~~~~~~~~~~-64\d_{18}+3616\d_{19})\,, \label{ccon}\\
&{\rm or}& \nonumber \\
\b_7 &=&-\frac{197}{25920}-\frac{1}{9}(14\g_{10}+86\g_{11}+678\g_{12}+24\d_{15}-696\d_{16}-820\d_{17}\nonumber \\ &&~~~~~~~~~~~~~~~~~~~-864\d_{18}-59712\d_{19}) \nonumber \,.\\
\end{eqnarray}
\subsection*{Criticality of counterterm actions}
As shown in appendix C, the equations of motion following from our action written as $I=\int \sqrt{|g|} L\equiv \int \sqrt{|g|} M^{1/2}$ is given by:
\be
0=-g_{db}L-2\nabla^a \nabla_{(d} g_{b)e}\N^e_a+g_{be}\Box \N^e_d+g_{bd} \nabla^a \nabla_c \N^c_a+2 R_{dc}\N^c_b\,,
\ee
with $\N$ defined through $$\N= \frac{N}{2\sqrt{M}}\,.$$ Then since $\N$ vanishes at zeroth order, the linearized equations of motion read
\be
0=-2\bnabla^a \bnabla_{(d} \bar g_{b)e}\tilde \N^e_a+\bar g_{be}\bbox \tilde\N^e_d+\bar g_{bd} \bnabla^a \bnabla_c \tilde \N^c_a+2 \bar R_{dc}\tilde \N^c_b\,,
\ee
where $\bar{}$ denotes the background value while $\tilde{}$ denotes linearization. Furthermore, we will treat the critical point as a limiting case\footnote{A similar analysis for DBI gravity in 3-dimensions was carried out in \cite{mohsen2}.} of $\fin\rightarrow 1$ so that we can treat $M\sim (\fin-1)^{3/2}+O(h^2)$ to be a constant. Then the problem boils down to working out the linearization of $N$. It is straightforward to see that the linearization of $N$ must take the form
\be
\tilde N_{ab}=c_1 \R^{L}_{ab}+c_2 \R^L \bar g_{ab}+c_3 h_{ab}\,.
\ee
For definiteness, let us consider $n=5$.
Imposing
\be
\bnabla^a h_{a b}=\bnabla_b h\,,
\ee
we simplify
\be
\tilde N_{ab}=\frac{c_1}{2}(\bnabla_a\bnabla_b h-\bbox h_{ab}-\frac{10}{L^2}h_{ab})+\frac{2}{L^2}(c_1+4c_2)h \bar g_{ab}+c_3 h_{ab}\,.
\ee
Using this we have
\be
\tilde N^a_a= (5 c_2+\frac{L^2}{4}c_3)\R^L\,.
\ee
Using a computer program (or otherwise) we can check that $c_3=\frac{4}{L^2}c_1$, $c_2=-\frac{c_1}{5}$ so that $5 c_2+\frac{L^2}{4}c_3=0$ and hence $\tilde N_a^a=0$. Here $c_1$ is given by
\be
c_1=-120 L^4  (\beta_7-\b_c)\,,
\ee
with $\b_c$ defined in eq.(\ref{ccon}). Using the identities in the appendix, we can show that the trace of the equations of motion leads to
\be
(-\frac{3}{2}c_1+8 c_2+L^2 c_3)\bbox \R^L-\frac{8}{L^2}(5 c_2+\frac{L^2 c_3}{4})\R^L=0\,.
\ee
So we have
\be
\bbox h=0\,.
\ee
In appendix (\ref{nogauge}) we have shown that $h$ cannot be gauged away.
After some tedious algebra we also find
\be
c_1\left(-\frac{32}{5}L^2\bnabla_a\bnabla_b h+\frac{2}{5}h\bar g_{ab}-L^4 \frac{1}{2}(\bbox+\frac{2}{L^2})(\bbox+\frac{2}{L^2})h_{ab}\right)=0\,.
\ee
If we separate out the trace bit from $h_{ab}$ by writing $\hat h_{ab}=h_{ab}-\frac{h}{5}\bar g_{ab}+\frac{1}{5}L^2 \bnabla_a \bar \nabla_b h$ and use $\bbox h=0$ then we get
\be
\frac{c_1}{2}(\bbox+\frac{2}{L^2})(\bbox+\frac{2}{L^2}){\hat h}_{ab}=0\,.
\ee
In a similar manner, we can repeat the above exercise for the square-root action for arbitrary dimensions. We have verified explicitly that for $d=3,5,7$,
\be
c_2=-\frac{c_1}{d},\quad c_3=\frac{d-1}{L^2}c_1
\ee
$c_1$ being a linear combination of parameters such that $c_1=0$ leads to a vanishing Euler anomaly.
This gives us
\be
\frac{(d-2)^2(d-1)}{2d}c_1 \bbox h=0\,,
\ee
and
\be
c_1\left(-2\frac{(d-1)^2}{d}L^2\bnabla_a\bnabla_b h+\frac{2}{d}h\bar g_{ab}-L^4 \frac{1}{2}(\bbox+\frac{2}{L^2})(\bbox+\frac{2}{L^2})h_{ab}\right)=0\,.
\ee
Writing  $$\hat h_{ab}=h_{ab}-\frac{h}{d}\bar g_{ab}+\frac{1}{d}L^2 \bnabla_a \bar \nabla_b h$$ we have as before
\be
\frac{c_1}{2}(\bbox+\frac{2}{L^2})(\bbox+\frac{2}{L^2}){\hat h}_{ab}=0\,.
\ee
The field $\hat h_{ab}$ is both transverse and traceless, i.e., $\bnabla^a \hat h_{ab}=0=\hat h$. Hence we are left with a transverse traceless graviton and a massless scalar. In terms of $\hat h_{ab}$ the linearized equations of motion are identical to what follows from critical gravity. Unlike the quadratic critical gravity theory where we do not have a propagating scalar, in our square-root theories there is the possibility of having a propagating scalar.

\section{Connections with log-CFTs}
Logarithmic CFTs were first introduced by Gurarie \cite{gurarie} in the context of 1+1d CFTs. Their connections with disorder CFTs, turbulence, quenched ferromagnets etc. \cite{loglects} make it worthwhile to construct AdS/CFT duals for studying them.
The connection between critical gravity in 3-dimensions and log-CFTs has been explored in detail by Grumiller, Johansson and collaborators \cite{gru, banom}.

As we have found above, the linearized equations of motion resemble those of critical gravity except for the propagating scalar mode. Moreover, we have found that in all the square-root theories, the Euler anomaly vanishes. Vanishing of a central charge is a smoking gun for underlying logarithmic CFTs. While a particular central charge vanishes, one also gets new central charges, for example the so-called ``b-anomaly" \cite{gurarie,banom}. In our examples, the $b$-anomaly is proportional to $c_1$. Vanishing of $c_1$ (e.g., the DBI counterterm and its generalizations) would lead to vanishing $b$-anomaly.

Establishing a precise relation with logarithmic CFTs requires calculation of stress tensor correlation functions which is beyond the scope of this paper. However, we point out one key fact which renders credence to the dual CFT being a logarithmic CFT.

As we have mentioned above, in logarithmic CFTs, the stress tensor satisfies $\langle T_{ab} T_{cd} \rangle\propto c \langle 0|0\rangle =0$. This would mean that either $c=0$ or that the vacuum satisfies $\langle 0|0\rangle=0$ or both. In the context of log-CFTs this implies that the vacuum should have a log-partner. In our $n=5$ model above, we have found that there are putative log-CFTs with $c\neq 0$ but $a=0$. This would mean that in these models we should have a degenerate vacuum with a log-partner. There is a very good candidate for this log-partner. It has been shown in \cite{teklog, mohsen} that there are {\it exact} AdS-wave solutions to quadratic gravity\footnote{See e.g., \cite{tekinri, ghodsi} for more exact solutions in the 3d case. } . These waves are given by
\be
ds^2=\bar g_{ab}dx^a dx^b+ h_{ab}dx^a dx^b=\frac{L^2}{z^2}\left[dz^2+(2 dx^+ dx^-+d{\bf x}^2_{d-3})\right]+2 V(x^+, z)(dx^+)^2\,,
\ee
where
\be
V(x^+,z)=v_1(x^+)z^{d-3}+\frac{v_2(x^+)}{z^2}+\frac{1}{d-1}(v_3(x^+)z^{d-3}+\frac{v_4(x^+)}{z^2})\log \frac{z}{L}\,.
\ee
Note that $h_{ab}$ in this context is not a small fluctuation. The criticality of the quadratic theory is crucial to have the log term in the solution.
Using the methods in \cite{gibbonsexact, teklog} it can be easily verified that these AdS-waves are solutions to our square-root theory. Furthermore, a few lines of algebra leads to the conclusion that these waves have vanishing action. In order to be a non-trivial solution, rather than just being asymptotically AdS, one needs to have $v_4\neq 0$. It is this solution that we propose as the log-partner to the usual AdS-vacuum. It would be interesting to see if there are yet more general solutions which do not have $h^a_a=0$ but rather $\Box h^a_a=0$ which are exact solutions.

\section{Discussion}

In this paper we studied AdS counterterms in the context of odd $d$-dimensional CFTs. For such CFTs, our counterterms led to a cut-off independent free energy on an $S^d$. It will be interesting to understand the implications for the conjecture relating this free energy to a c-function \cite{mch,klebaF,mscarb}. It should be noted that there now appears to be a proof of the c-theorem for 4-dimensional CFTs \cite{zohar}, which possibly generalizes to even dimensions, while in odd dimensions a similar proof is lacking \cite{ms1,mscarb,klebaF}.

We have worked with the metric formalism of gravity for the question of
counterterms. However, if one worked in the first order Palatini
formalism of AdS gravity in even dimensions then it is known that by
adding the topological Euler density term with a specific coefficient one
can make the action vanish identically for any locally AdS spacetime
\cite{acotz1, acotz2}. Such an action will be manifestly cutoff
independent. It will be interesting to see if these two approaches can be
related.

When we treated the counterterm action as an independent action for gravity, we found that the linearized equations of motion for the gravity were similar to what has been found in the context of critical gravity in recent times. Interestingly\footnote{We thank Daniel Grumiller for pointing this out to us.}, even the Chern-Simons term in 3d can arise as a surface term from a 4d gravity theory which includes the Chern-Pontryagin density \cite{mcnees}. Thus, curiously, all interesting higher derivative theories considered in recent times related to a critical point, appear to be related in some way to the boundary of AdS space.
Although at the linearized level our theories are very similar to the quadratic critical gravity theories in the literature and contain the usual log-modes, we also find a propagating massless scalar. As we pointed out in section 2, one can get a propagating scalar in quadratic critical gravity with negative Euler anomaly. In our case, when $I_{ct}$ was considered as a toy model for AdS/log-CFT, we found that the Euler anomaly was always zero while the other central charges could be positive depending on the choice of parameters. Since the Schwarzschild black hole entropy in the critical theories happen to vanish, it will be interesting to see what happens to the holographic entanglement entropy on a sphere. Even though the theories are non-unitary, the entanglement entropy must be non-vanishing. It will also be interesting to see if the counterterms needed in Lovelock theories \cite{ayale} can lead to critical gravity along the lines of this paper.

We found ``square-root" actions which contain an infinite set of specific higher derivative corrections. As we mentioned earlier, setting $\lambda=-1/24$ in eq. (3.4) leads to the DBI gravity theory \cite{desergibbons} \footnote{Without the $\R_{ab}$, this is the Mann-Lau counterterm \cite{mann,lau}.} in 3-dimensions \cite{jatkarsinha}:
\begin{equation} \label{eqm1}
I_{ct}=-\frac{2L^2}{\ell_P^2} \sqrt{-{\rm det}(\R_{ab}-\frac{1}{2}\R
  \g_{ab}-\frac{1}{L^2} \g_{ab})}\,.
\end{equation}
A question naturally arises if there is a choice of parameters for the counterterms in AdS$_6$ which leads to a simple action.
It turns out that for a special choice of parameters corresponding to
\begin{eqnarray}
\b_7 &=&\frac{1}{576}\,,\quad \gamma_{10}=\gamma_{11}=\gamma_{12}=0\,,\\
\d_{15}&=&\frac{2}{27}\,,\quad \d_{16}=-\frac{5}{108}\,,\quad \d_{17}=\frac{17}{864}\,,\quad \d_{18}=-\frac{1}{24}\,,\quad \d_{19}=-\frac{53}{34560}\,,
\end{eqnarray}
the square-root action can be written in a very compact way.
Defining
\begin{eqnarray}
G_{ab}&=&\R_{ab}-\frac{1}{4}\R \gamma_{ab}-\frac{1}{L^2}\gamma_{ab}\,,\\
H_{ab}&=&\R_{ab}-\frac{1}{4}\R \g_{ab}+\frac{3}{2L^2}\g_{ab}\,,
\end{eqnarray}
the counterterm action which cancels off all the cut-off for $S^5$ and $R\times S^4$  can be written as
\be
I_{ct}=-\frac{L^4}{\lp^4}\left(-\frac{2^3}{5\times 3^3} \epsilon^{a_1 b_1 c_1 d_1 e_1}\epsilon^{a_2 b_2 c_2 d_2 e_2}G_{a_1 a_2}G_{b_1 b_2}G_{c_1 c_2}H_{d_1 d_2}H_{e_1 e_2}\right)^{1/2}\,.
\ee
The corresponding stand-alone theory has $c=a=0$. The above form of the action makes it clear that fluctuations around AdS will begin at $O(h^3)$ since there are 3 powers of $G$ which vanishes onshell. This is very similar to the DBI case in \cite{jatkarsinha}. Since there are no gravity waves around AdS, the theory is non-dynamical. Had  $H$ and $G$ featuring above been the same, the theory would be again a DBI gravity theory. However, what arises is rather a generalization of DBI. It will be interesting to see if there is a way to rewrite this theory in terms of a gauge field as in the 3 dimensional case studied in \cite{jatkarsinha}. The above results hint at a way to extend the DBI form of the counterterm for arbitrary (odd) dimensions. First we define
\be
G^{(i)}=\R_{ab}-\frac{1}{d-1}\R \g_{ab}+\frac{\lambda_i}{L^2}\g_{ab}\,.
\ee
For $d=3,5$ above we observe that $\lambda_i$'s are the roots of the equation $f(\lambda=\frac{L^2}{\Lambda^2})|_{\partial M=S^d}=(I_{bulk}+I_{GH})^2=0$. In other words we compute the on shell action for $AdS_{d+1}$ when the boundary is $S^d$, write it as a function of $\lambda=L^2/\Lambda^2$ and work out the zeros of its square which will give the $\lambda_i$'s. Then the counterterm in arbitrary dimensions is
\be\label{gencd}
I_{ct}=-(d-1)\frac{L^{d-1}}{\lp^{d-1}}\left(-\frac{1}{\prod_i \lambda_i}\frac{1}{d!} \epsilon^{a_1\cdots a_d}\epsilon^{b_1 \cdots b_d} \prod_{i=1}^d G^{(i)}_{a_i b_i}\right)^{1/2}\,.
\ee
We have checked that this is indeed true for $d=3,5,7$. In each case there are always 3 roots with $\lambda_i=1$.

We studied cases where the bulk theory was even dimensional so that the counterterm action was odd dimensional. What about odd dimensional bulk? In this case, there would be a conformal anomaly in the boundary CFT which would require the addition of non-local counterterms or local counterterms with additional fields. It is for this reason we have refrained from addressing this interesting case. We have checked that when $D=5,d=4$ and $D=7,d=6$, it is possible to write a counterterm which gets rid of all the cutoff dependence in the Euclidean on-shell action for boundary topologies $S^d, R \times S^{d-1}$ arising from the log-independent terms. There is a DBI way of writing this counterterm that follows from eq.(\ref{gencd}). Consider $D=5,d=4$. In this case there are two roots with $\lambda_i=-2$, one with $\lambda_i=-1$ and one infinite root. The infinite root is not a problem since this essentially means that we replace one of the $G_i$'s with $1/L^2 \g_{ab}$ and we still get a finite result due to the $1/\prod_i \lambda_i$ factor. Thus eq.(\ref{gencd}) seems to cover all cases including even dimensions! With the above choice of counterterms we would get the Casimir energy on $R\times S^{d-1}$ to vanish in even dimensions. This is consistent with the fact that our choice of counterterm would lead to ambiguities \cite{bk}, eg, a $\Box R$ term in the trace anomaly \cite{bk} in $d=4$ so that a comparison between the gauge and gravity Casimir energies can only be made after matching the coefficients of the $\Box R$ term in the trace anomaly. A more quantitative verification of this statement would be gratifying. Curiously, if one considered the results in \cite{robcounter} for even dimensional CFTs and considered the finite contributions from subleading counterterms (which were dropped in their analysis), one would conclude that the Casimir energy is actually zero. This is again consistent with the fact that the subleading counterterms would generate terms analogous to $\Box R$ in the trace anomaly, which need to be matched between the two sides before any comparison is attempted.

Another question that we have not addressed is the following. In the context of $I=I_{bulk}+I_{surf}+I_{ct}$ does it make sense to consider equations of motion for the boundary metric? It is certainly true that if we considered Dirichlet boundary conditions then we set $\delta g=0$ at the boundary so that one does not get corrections to the bulk equations of motion from the boundary. This was also the reason why one needs the Gibbons-Hawking term. The case where one considers Neumann boundary conditions is murkier. It has been claimed that it is possible to set the ``boundary free" in \cite{compere}. It is possible that in this context it is sensible to consider the equations of motion for the boundary metric separately from the bulk. However, we leave a more accurate analysis of this question for future work. Naively, it seems that if we considered the limit $\lp\rightarrow \infty, L\rightarrow \infty$ keeping $L^{d-1}/\lp^{d-1}$ fixed, we could decouple the counterterm action from the bulk action. For the counterterm action, this sets the $\lambda_i$'s in eq.(\ref{gencd}) to zero. This would be similar to a ``flat-space" limit. However, taking $\lp\rightarrow\infty$ would make quantum corrections in the bulk very important so it may not be a valid limit (it also is reminiscent of the tensionless limit in string theory). It will be interesting to examine the graviton dynamics ala DGP \cite{dgp} in this context when no such decoupling is attempted.

We would like to make some speculative comments about connection between the counterterm actions and singletons. As pointed out in \cite{flatofronsdal}, the singleton equation of motion reads:
\be
(\bbox+\#/L^2)^2 \phi=0\,,
\ee
where $\#$ is some dimension dependent constant. The fact that the linearized equations of motion for the metric look similar makes one wonder if there is a connection between the counterterm action and singletons\footnote{The connection between log-CFTs and singletons was already pointed out in \cite{kogans}.}. The following observation makes such a connection more poignant.
The AdS$_4$ DBI counterterm can be rewritten in terms of a gauge field  \cite{jatkarsinha}. The gauge field in question is the SL(2,R) gauge field that features in rewriting 3d Einstein gravity as a Chern-Simons theory, namely
$
  A^{a\pm} = \omega^a \pm \frac{1}{\ell} e^a\, ,
$ where $\omega^a = \epsilon^{abc}\omega_{bc}/2$ is the dualised spin
connection and $e$ is dreibein and $a$ is a gauge index corresponding
to SL(2,R). In terms of the field strength $\mathcal{F}^a = \frac{1}{2} (F^{a+} +
  F^{a-})$, one finds that
  \begin{equation}
  \label{eq:11}
  I_{ct} \propto \sqrt{\det{\star \mathcal{F}^{a\mu}}}\, .
\end{equation}
However, $\star \mathcal{F}^{a\mu}=\partial^\mu \Phi^a$ where $\Phi^a$ is some scalar carrying an SL(2,R) index and as such the action can be rewritten in terms of a scalar field. It is tempting to think that $\Phi^a$ is related to the AdS$_4$ singleton field.

Finally it will be worthwhile to use our counterterms to probe holographic renormalization group flows  \cite{holorg, ranga}. As was pointed out in \cite{ranga} the total action is written as
$ S=\int_{r<1/\epsilon} d^{d+1}x \sqrt{g} {\mathcal L}+
S_B$ where $S_B$ is a boundary action defined on $r=1/\epsilon$ and
can be viewed as a boundary state for the bulk theory in the region
$r<1/\epsilon$. One fixes $S_B$ by demanding $\partial_\epsilon S=0$. This is related to the problem we have solved, except that our counterterms for $d\geq 5$ only work for boundaries with vanishing Weyl tensor while a general solution will require counterterms to work for any topology. We have attempted to remedy this shortcoming by considering Weyl$^n$ corrections making results for $H^2 \times S^3$ type boundaries cut-off independent in $d=5$, leaving a general study for future work. Nonetheless, it will be worthwhile computing stress tensor correlation functions using our counterterm actions and making the connection with  \cite{ranga} more concrete.

\section*{Acknowledgments}{We thank Alex Buchel, Janet Hung, Dileep Jatkar, Robert Leigh, Robert Mann, Robert Myers and Miguel Paulos  for useful discussions. AS thanks Daniel Grumiller and Zohar Komargodski for useful correspondence. Special thanks to Robert Myers for going through the draft in detail and for useful comments. AS thanks Perimeter Institute for hosting him during the course of this work and the University of Cincinnati for hospitality where part of this work was presented in the November-2011 SPOCK meeting. }
\appendix
\section{A simple method to compute holographic Weyl anomalies in 4d CFTs}\label{simpan}
Four dimensional CFT's are characterized by two central charges $c$ and $a$. These are defined through
\be
\langle T_{a}^{\ a}\rangle= \frac{c}{16\pi^2} I_4 -\frac{a}{16\pi^2} E_4\,,
\ee
where
\be
I_4= R_{a b c d} R^{a b c d}-2 R_{ab}R^{ab} +\frac{1}{3} R^2\,,\quad E_4= R_{a b c d} R^{a b c d}-4 R_{ab} R^{ab} +R^2\,.
\ee
In order to compute $c,a$ we follow the procedure described in  \cite{renorm1}. Here we start with the gravity action and use Fefferman-Graham expansion
\be\label{fg}
ds^2=\frac{\tilde L^2}{4 \r^2} d\r^2+ \frac{g_{i j}}{\rho} dx^i dx^j\,,
\ee
where
\be
g_{i j}=g_{(0) i j}+ \rho g_{(1)i j}+\rho^2 g_{(2) ij}+ \cdots \,,
\ee
with $g_{(0)}$ denoting the boundary metric. The procedure in  \cite{renorm1, two}  instructs us to plug in this expansion into the gravity action. On-shell $g_{(2)}$ drops out and we are left with an action involving $g_{(0)}$ and $g_{(1)}$.  To extract the conformal anomaly, we focus on terms leading to a log divergence. This leads to
\begin{eqnarray} \label{holact}
S_{ln}&=&{\cal N} \int d^4 x \sqrt{g_{(0)}} \bigg{[} \bigg{(} t_1 r^{(0)^2}+t_2 {\rm ric}^{(0)^2}+t_3 {\rm rim}^{(0)^2}\bigg{)}\nonumber \\
      &+& A r^{(0)ij} g_{(1)ij}+B r^{(0)} {\rm tr~} g_{(1)} + C {\rm tr} g_{(1)}^2+D ({\rm tr} g_{(1)})^2\bigg{]}\,,
\end{eqnarray}
where $r^{(0)}$ is the Ricci scalar, ${\rm ric}$ is the Ricci tensor and ${\rm rim}$ is the Riemann tensor all made from $g_{(0)}$.
We solve $g_{(1)}$ w.r.t $g_{(0)}$. This leads to
\be
S_{ln}=-\frac{1}{2} \int d^4 x \sqrt{g_{(0)}} {\cal T}\,,
\ee
where we identify
\be
{\cal T}=\langle T_{a}^{\ a}\rangle= \frac{c}{16\pi^2} I_4 -\frac{a}{16\pi^2} E_4\,.
\ee

This procedure although conceptually straightforward gets messy as one considers higher derivative terms in the gravity theory. Here we outline a simple method to compute the anomaly coefficients. We will choose for $g_{ij}$ the metric
\be \label{ansatz}
ds^2= u[1+\alpha\rho] (-R^2 dt^2+ \frac{dR^2}{u R^2})+v[1+\beta\rho](d\theta^2 +\sin \theta^2 d\phi^2)\,,
\ee
i.e., it is of the form AdS$_2 \times$ S$^2$.  Plugging this back into the lagrangian, we will extract the coefficient of the $1/\rho$ term which we will call $L_{\ln}$. Using this we work out the equations of motion for $\alpha,\beta$ which are simply given by
\be
\partial_\alpha L_{\ln}=0\,,\qquad \partial_\beta L_{\ln}=0\,.
\ee
Next we compute the four dimensional $I_4, E_4$ made from $g_0$ which are given by
\be
I_4=\frac{4 (u-v)^2}{3 u^2 v^2}\,,\qquad E_4=-\frac{8}{uv}\,.
\ee
Using these it is straightforward to show that
\begin{eqnarray}
c&=&\lim_{v\to\infty} \frac{48\pi^2}{\sqrt{g_0}} L_{\ln}{\bigg |}_{u=1}\,,\\
a&=& \lim_{v\to 1} \frac{8 \pi^2}{\sqrt{g_0}} L_{\ln}{\bigg |}_{u=1}\,.
\end{eqnarray}
One can use these formulae to check $c,a$ in the literature, e.g.,  \cite{quasi}.
\section{Useful identities}
We will follow the conventions in  \cite{wald}. The following identities are useful in our analysis. We consider Einstein spaces of the form
\be
\R_{abcd}=\frac{2\Lambda}{(n-1)(n-2)}(g_{ac}g_{bd}-g_{ad}g_{bc})\,.
\ee
We will frequently be using the gauge choice $\bnabla^a h_{ab}=\bnabla_b h$. 
We define the linearized Ricci tensor by $\R^L_{ab}$ while $\R^L$ is the linearized Ricci scalar.
\begin{eqnarray}
\R^{L}_{ab}&=& \bnabla^c \bnabla_{(a} h_{b)c}-\frac{1}{2}\bbox h_{ab}-\frac{1}{2}\bnabla_a\bnabla_b h\,, \label{Rlab}\\
\R^L&=& \bnabla^a \bnabla^b h_{ab}-\bbox h-\frac{2\Lambda}{n-2}h\,,\\
\bnabla_a\bnabla_b h_{cd}&=& \bnabla_b\bnabla_a h_{cd}+\bar \R_{abd}^{\ \ \ e}h_{ce}+\bar \R_{abc}^{\ \ \ e}h_{de}\,, \label{use1}\\
\bnabla^b \R^L_{ba}&=&-\frac{1}{2}\bnabla_a \R^L\,,\\
\bar g^{ab} \R^L_{ab}&=&0\,,\\
\bbox \bnabla_a h &=& \bnabla_a\bbox h+\frac{2\Lambda}{n-2}\bnabla_a h\,,\\
\bbox \bnabla_a\bnabla_b h &=& \bnabla_a\bnabla_b \bbox h+\frac{4n \Lambda}{(n-1)(n-2)}\bnabla_a\bnabla_b h-\frac{4\Lambda}{(n-1)(n-2)}\bar g_{ab}\bbox h\,,\\
\bnabla^a\bbox h_{ab}&=&\bbox\bnabla^a h_{ab}+\frac{2\Lambda}{n-2}\bnabla_b h\,.
\end{eqnarray}
The derivation of the fourth identity is as follows. We start with eq.(\ref{Rlab}). Using eq.(\ref{use1}), we have
\begin{eqnarray}
\R^L_{ab}&=&\frac{1}{2}\bnabla_a\bnabla_b h+\frac{2\Lambda n}{(n-1)(n-2)}h_{ab}-\frac{2\Lambda}{(n-1)(n-2)}h \bar g_{ab}-\frac{1}{2}\bbox h_{ab}\,,\\
\R^L &=& -\frac{2\Lambda}{n-2}h\,.
\end{eqnarray}
Then we have
\begin{eqnarray}
\bnabla^b \R^L_{ba}&=&\frac{1}{2}\bnabla^b\bnabla_a\bnabla_b h+\frac{2\Lambda n}{(n-1)(n-2)}\bnabla^b h_{ab}-\frac{2\Lambda}{(n-1)(n-2)} \bar g_{ab}\bnabla^b h-\frac{1}{2}\bnabla^b \bbox h_{ab} \nonumber\\
 &=& \frac{\Lambda}{n-2}\bnabla_a h \\
 &=& -\frac{1}{2}\bnabla_a \R^L\,.
 \end{eqnarray}
\section{Equations of motion}
We have an action that is of the form
\be
S=\int \sqrt{|g|} L( R_a^b)\,,
\ee
i.e., we can consider terms like $R=R_a^b \delta^a_b,  R_{ab}R^{ab}=R_a^b R_b^a$ etc and the contraction of indices is being done with the Kronecker Delta. The advantage of this way of writing is that we do not have to worry about varying the metric independently in $L$. The variation of $S$ leads to
\be
\delta S=\int \delta(\sqrt{|g|}) L+ \sqrt{|g|} \frac{\partial L}{\partial R^a_b} \delta R^a_b\,.
\ee
Now we can use
\be
\delta R^d_b=g^{da}(\nabla_c \delta \Gamma^c_{ab}-\nabla_b \delta \Gamma^c_{ac})+R_{ab}\delta g^{da}\,,
\ee
to get
\be
\delta S=\int \frac{1}{2}\sqrt{|g|}(-g_{db}L-2\nabla^a \nabla_{(d} g_{b)e}\frac{\partial L}{\partial R^a_e}+g_{be}\nabla^2 \frac{\partial L}{\partial R^d_e}+g_{bd} \nabla^a \nabla_c \frac{\partial L}{\partial R^a_c}+2 R_{dc}\frac{\partial L}{\partial R^b_c})\delta g^{bd}\,.
\ee
This should be the starting point to program a computer to evaluate the equations of motion starting from an arbitrary $L(R_a^b)$. In our case the basic building blocks are the following:
\begin{eqnarray}
\Q_0&\equiv& \frac{\partial \R}{\partial \R^a_b}=\delta_a^b\,,\quad \Q_1\equiv\frac{\partial \R_2}{\partial \R^a_b}=2 \R_a^b\,,\quad \Q_2\equiv\frac{\partial \R_3}{\partial \R^a_b}=3 \R_a^c \R_c^b\,,\\ \Q_3&\equiv&\frac{\partial \R_4}{\partial \R^a_b}=4 \R_a^c \R_c^d \R_d^b\,,\quad \Q_4\equiv\frac{\partial \R_5}{\partial \R^a_b}=5\R_a^c \R_c^d \R_d^e \R_e^b\,.
\end{eqnarray}
Using these it should now be straightforward to work out the equations of motion.
For instance, writing $L=M^{1/2}$, we will need to work out $\partial M/\partial \R^a_b=N_a^b$. For the square-root action in $d=5$, we find that $N$ is given by
\begin{eqnarray}
N&=&\a_2 L^2 \Q_0+ L^4(2\a_3 \R \Q_0+\a_4 \Q_1)+L^6[\b_5 \Q_2+\b_6(\R \Q_1+\R_2\Q_0)+3\b_7 \R^2 \Q_0]\nonumber  \\
 &+& L^8 [\g_8 \Q_3+\g_9(\R \Q_2+\R_3 \Q_0)+2\g_{10}\R_2 \Q_1+\g_{11}(\R^2 \Q_1+2 \R_2 \R \Q_0)+4 \g_{12}\R^3 \Q_0]\nonumber \\
 &+& L^{10}[\d_{13}\Q_4+\d_{14}(\R \Q_3+\R_4 \Q_0)+\d_{15}(\R_2\Q_2+\R_3\Q_1)+\d_{16}(\R^2 \Q_2+2\R_3 \R \Q_0) \nonumber \\
 &&~~~~~~~~~~+\d_{17}(\R^3 \Q_1+3 \R_2 \R^2 \Q_0)+\d_{18}(2\R_2 \R \Q_1+\R_2^2 \Q_0)+5 \d_{19} \R^4 \Q_0]\,.
 \end{eqnarray}
Now defining
\be \label{calN}\N= \frac{N}{2\sqrt{M}}\,,\ee
the equations of motion can be written as
\be
0=-g_{db}L-2\nabla^a \nabla_{(d} g_{b)e}\N^e_a+g_{be}\Box \N^e_d+g_{bd} \nabla^a \nabla_c \N^c_a+2 R_{dc}\N^c_b\,.
\ee
So in order to program a computer, we simply need to define the two index symmetric tensor $\N$ and compute the covariant derivatives in the usual way.

\section{Proof that $\gamma$ cannot be gauged away} \label{nogauge}
We want to start with
\be\label{ap1}
\bnabla^a h_{ab}=\bnabla_b h\,,
\ee
and consider restricted gauge transformations
\be\label{tra}
\tilde h=h_{ab}+\bnabla_a w_b+\bnabla_b w_a \implies \tilde h=h+2\bnabla^a w_a\,,
\ee
such that
\be
\bnabla^a \tilde h_{ab}=\bnabla_b\tilde h\,,
\ee
and the equations of motion are respected. We wish to see if there exists $w_a$ such that $\tilde h=0$ starting with a non-zero $h$.
Plugging eq.(\ref{tra}) into eq.(\ref{ap1}) we need
\be
\bbox w_b-\bnabla_b\bnabla^a w_a+\bar \R_b^c w_c=0\,.
\ee
which leads to
\be \bbox \nabla^a w_a=\bnabla^a \bbox w_a -\frac{n-1}{L^2}\bnabla^a w_a\,,
\ee
whereas using
\be
\bbox \bnabla_a w_b=\bnabla_a \bbox w_b+\frac{2}{L^2}\bar g_{ab}\bnabla^c w_c-\frac{2}{L^2}\bnabla_b w_a-\frac{n-1}{L^2}\bnabla_a w_b\,,
\ee we have
\be \bbox \nabla^a w_a=\bnabla^a \bbox w_a +\frac{n-1}{L^2}\bnabla^a w_a\,,
\ee
so that
\be
\bnabla^a w_a=0\,.
\ee
In other words, the allowed restricted gauge transformations must be transverse and hence we cannot gauge $h$ away. This proves that $h$ can be dynamical.

\section{Constraints from simple c-theorems}
In  \cite{sinha1,sinhacos} higher derivative lagrangians were constrained by demanding the existence of a simple c-theorem. A recursive prescription to build higher order lagrangians in 3d based on the simple c-theorem constraint  \cite{sinha1} was given in  \cite{paulos}. Consider the $d=5$ lagrangians. On general grounds, since the theories we have are not unitary, such a c-theorem may not exist. Nonetheless we ask if there is a choice of parameters which allows for the existence of a c-function. We start by writing
\be
ds^2=e^{2A(r)}(-dt^2+dx_1^2+dx_2^2+dx_3^2)+dr^2\,,
\ee
and impose
\be
T_t^t-T_r^r\geq 0
\ee
using the null energy condition \footnote{See  \cite{naka} for an interesting connection between the null energy condition and the Zamolodchikov-Polchinski theorem.}. The question is if using the equations of motion we can find a $a(r)$ such that $a'(r)\geq 0$ using the null energy condition and such that at the fixed points $a(r^*)$ becomes a central charge of the theory. The strategy in  \cite{sinha1} was to choose parameters in higher derivative lagrangians such that $T_t^t-T_r^r$ did not have $A''', A''''$ terms once the equations of motion were used. Proceeding in a similar manner, we find that
\begin{eqnarray}
\b_7 &=& -\frac{7}{5184}\,, \g_{11}=-\frac{10}{13}\g_{10}\,, \g_{12}=\frac{\g_{10}}{13}\,,\\
\d_{16}&=&-\frac{83}{991440}-\frac{5}{17}\d_{15}+16\d_{19}\,,\\
\d_{17}&=&\frac{5}{528768}+\frac{\d_{15}}{17}-10\d_{19}\,,\\
\d_{18}&=&\frac{151}{15863040}-\frac{5}{17}\d_{15}+13\d_{19}\,,\\
\end{eqnarray}
leads to
\be\label{nec}
T_t^t-T_r^r=\frac{L^2 A''(r)}{4\sqrt{1-L^2 A'(r)^2}}\leq 0\,,
\ee
so that
\be
a(r)=\frac{1}{A'(r)}\sqrt{1-L^2 A'(r)^2}\,,
\ee
satisfies $a'(r)\geq 0$ once we use $A''(r)\leq 0$ which follows from eq.(\ref{nec}). The above choice of parameters leads to $a=0$ at the fixed point and is consistent with the fact that both $c,a$ vanish with the above choice of parameters. Our proposed c-function works also for $d=3$ and is different from the proposal in  \cite{tekinc} which only gives a monotonic function which is proportional to the central charge at the fixed point rather than being equal to it. We have been unable to argue for the existence of a c-function in the general case and one need not exist. Furthermore, unlike the $d=3$ case where the existence of a simple c-function led to the DBI gravity action, in $d=5$ we have been unable to rewrite the resulting action from the above choice of parameters in a DBI like manner.

\end{document}